\documentclass[aps,prd,onecolumn,nofootinbib,superscriptaddress]{revtex4}
\usepackage[colorlinks=true,linkcolor=blue,urlcolor=blue,filecolor=black,citecolor=red,pdfstartview=FitV,pdftitle={},pdfsubject={},pdfkeywords={},pdfpagemode=None,bookmarksopen=true]{hyperref}
\usepackage{graphicx}%Include figure files
\usepackage{amsmath}
\usepackage{amsfonts}
\usepackage{amssymb,ulem}
\usepackage{color}%
\usepackage{dcolumn}% Align table columns on decimal pointl
\usepackage{braket}
\usepackage{eurosym}
\usepackage[usenames,dvipsnames,svgnames]{xcolor}

\newcommand{\RNum}[1]{\uppercase\expandafter{\romannumeral #1\relax}}

\usepackage[title]{appendix}
\usepackage{wrapfig,boxedminipage,setspace,subfigure,epsfig}

\begin{document}
\baselineskip=0.5 cm
%%%%%%%

\title{Exploring black holes with multiple photon spheres by interferometric signatures}

\author{Xi-Jing Wang}
\email{xijingwang01@163.com}
\affiliation{School of Physics and Technology, Wuhan University, Wuhan 430072, China}

\author{Yuan Meng}
\email{mengyuanphy@163.com}
\affiliation{School of Physics and Astronomy, China West Normal University, Nanchong, 637002, China}

\author{Xiao-Mei Kuang}
\email{xmeikuang@yzu.edu.cn}
\thanks{corresponding author}
\affiliation{Center for Gravitation and Cosmology, College of Physical Science and Technology, Yangzhou University, Yangzhou, 225009, China}

\author{Kai Liao}
\email{liaokai@whu.edu.cn}
\thanks{corresponding author}
\affiliation{School of Physics and Technology, Wuhan University, Wuhan 430072, China}

\begin{abstract}

In this paper, we investigate the interferometric signatures of hairy Schwarzschild black holes (hSBHs) that have either single or double photon spheres. Our interest mainly stems from two considerations: (i) the photon ring structure in black hole images produces strong and universal interferometric signatures on long baselines, enabling precision measurements of black hole parameters and testing gravitational theory; (ii) the hSBH describes the deformation of standard Schwarzschild black hole (SBH) induced by additional sources, and they can feature double photon spheres within certain parameter regimes. Using both analytical and numerical methods, we find that for a hSBH with a single photon sphere, the complex visibility amplitude of the image exhibits damped oscillations. A similar behavior appears in the double photon sphere case when the inner photon sphere has lower effective potential than the outer one, as the photons near the inner photon sphere remain trapped by gravity. However, when the inner potential is higher, a beat pattern rises. Our findings reveal that the complex visibility amplitude can encode the signature of the photon sphere structure of the central black hole.

\end{abstract}
%%%%%%%

\maketitle

\tableofcontents
\newpage

\section{Introduction}

Recently, the Event Horizon Telescope (EHT) has achieved remarkable breakthroughs in high-resolution images of the supermassive black holes M87* \cite{EventHorizonTelescope:2019dse,EventHorizonTelescope:2019uob,EventHorizonTelescope:2019jan,EventHorizonTelescope:2019ths,EventHorizonTelescope:2019pgp,EventHorizonTelescope:2019ggy,EventHorizonTelescope:2021bee,EventHorizonTelescope:2021srq} and Sgr A* \cite{EventHorizonTelescope:2022wkp,EventHorizonTelescope:2022apq,EventHorizonTelescope:2022wok,EventHorizonTelescope:2022exc,EventHorizonTelescope:2022urf,EventHorizonTelescope:2022xqj}. These images confirm the existence of black holes predicted by general relativity (GR), and disclose their optical features where a bright ring surrounding the central dark region, usually referred to as the black hole shadow. The bright region originates from photons emitted by the accretion disk outside the black hole, which are gravitationally lensed by the black hole and eventually reach the observer. The black hole shadow and the observation data obtained by EHT, which are believed to encode the crucial information about spacetime geometry, provide a powerful tool for constraining black hole parameters \cite{Tsukamoto:2014tja,Kumar:2018ple,Wang:2017hjl,Wang:2018prk,Hu:2020usx,Afrin:2021imp,Tsukamoto:2021caq,Meng:2022kjs,Li:2020drn,Kuang:2022ojj,Meng:2023wgi,Cao:2024vtq,Chen:2022kzv,Xavier:2023exm,Kouniatalis:2025itj}, placing bound on the size of extra dimensions \cite{Vagnozzi:2019apd,Banerjee:2019nnj,Tang:2022hsu} and testing GR or alternative theories of gravity \cite{Cunha:2015yba,Mizuno:2018lxz,Psaltis:2018xkc,Khodadi:2021gbc,Vagnozzi:2022moj,Zhang:2023okw,Kuang:2024ugn,Yang:2024utv,Wang:2024lte,Yue:2025fly}, and so on.

In the Universe, astrophysical black holes are typically surrounded by emitting materials, such as an accretion flow of hot, radiating gas, which governs their observed appearances \cite{Yuan:2014gma}. While it is difficult to mimic the astrophysical process between black hole and accretion flow from theoretical study, general relativistic magnetohydrodynamics enables first-principle simulations that ultimately allows us to reconstruct event-horizon-scale images of the black holes \cite{EventHorizonTelescope:2019dse,EventHorizonTelescope:2019pcy}. Nevertheless, the main characteristics of black hole images can also be captured by some simplified accretion models, with the use of much less resources. In \cite{Gralla:2019xty}, the authors assumed that the SBH is illuminated by an optically and geometrically thin accretion disk. They classified the structural composition of black hole images into three classes by the number of times ($m$) light intersecting with the accretion disk: direct ($m=1$, primary image), lensed ring ($m=2$, secondary image) and photon ring ($m\geq 3$, higher-order image) emissions. Their results demonstrated that the direct emission dominates observed intensity, with lensed ring emission secondary and photon ring emission negligible. Inspired by their studies, the photon ring and observational appearances of black holes surrounded by accretion disks have been extensively investigated in modified theories of gravity \cite{Zeng:2020vsj,Peng:2020wun,Bacchini:2021fig,Okyay:2021nnh,Guo:2021bhr,Boshkayev:2022vlv,Wen:2022hkv,Hou:2022eev,Chakhchi:2022fls,Wang:2022yvi,Wang:2023vcv,Uniyal:2023inx,Zhang:2024hix,Chen:2025ifv} and references therein.

The bright ring in the black hole image comprises an infinite sequence of self-similar subrings, each formed by photons orbiting the black hole multiple times \cite{Luminet:1979nyg}. These subrings asymptotically approach the edge of black hole shadow, known as the critical curve \cite{Gralla:2019xty}. As the index $m$ increases, the subring exhibits exponentially narrower and weaker, with seemingly negligible contributions from high-order subrings to the observed intensity. A natural issue is whether these subrings can be resolved by the EHT detectors, the study of which is very important to measure more precise ring structure and the dynamics of central supermassive black hole. Recently, the authors of \cite{Johnson:2019ljv} made the first attempt to analyze this issue. They found that these subrings could produce strong and universal signatures on long interferometric baselines such that high-frequency ground array or low Earth orbits could feasibly measure the lensed ring of M87 and Sgr A*, while the stations on the Moon and at the second Sun-Earth Lagrange point might resolve the first and second photon subrings, respectively. Though the subrings are exponentially thinner and dimmer as they approach the critical curve, they produce a cascade of damped oscillations in the complex visibility domain that reflect the properties of black hole metric. This will enable precision measurements of black hole parameters and tests of GR through interferometric signatures in the future \cite{Johnson:2019ljv,Aratore:2021usi}. It is noteworthy that studies in the fields such as galactic dynamics and gravitational wave have provided important insights into probing black holes and their surrounding environments. Please refer to \cite{Cardoso:2021wlq,Cardoso:2022whc,Destounis:2022obl,Chen:2024nua} for more details.

The no-hair theorem in GR states that the classical black holes are only described by mass, angular momentum, and electric charge \cite{Ruffini:1971bza}. However the black hole may carry hair by introducing additional field such as scalar field \cite{Herdeiro:2015waa}. Recently, a hSBH generated by gravitational decoupling (GD) approach has been proposed without considering specific fundamental fields \cite{Ovalle:2020kpd,Contreras:2021yxe}. The hSBH characterizes the deformations of known solutions in GR, which arise from additional sources in the energy-momentum tensor. Subsequently, the physical properties of this black hole and its rotating counterpart have been investigated extensively, such as thermodynamics \cite{Mahapatra:2022xea}, quasinormal modes \cite{Cavalcanti:2022cga,Yang:2022ifo,Li:2022hkq}, strong gravitational lensing, black hole shadow and image \cite{Islam:2021dyk,Afrin:2021imp,Meng:2024puu,Li:2025ixk,Meng:2025ivb}, Lense-Thirring effect \cite{Wu:2023wld}, and gravitational waves from extreme mass ratio inspirals \cite{Zi:2023omh}. In particular, the gravitational waves from such extreme mass ratio inspirals are among the primary targets for future space-based gravitational wave detectors, such as the Laser Interferometer Space Antenna (LISA) \cite{LISA:2017pwj,Barausse:2020rsu,LISA:2022yao,LISA:2022kgy,Cardenas-Avendano:2024mqp,Gair:2012nm}.

It is noteworthy that, in contrast to SBH with a single photon sphere outside the event horizon, the hSBH admit an additional photon sphere in certain parameter regimes \cite{Guo:2022ghl}. These multiple photon spheres bring in richer and distinctive observational signatures. In particular, the presence of additional photon sphere has been shown to modify the black hole images \cite{Meng:2023htc}, and also found to closely associated with the echo signals of the probe scalar field \cite{Yang:2024rms} \footnote{While our focus is on the particular hSBH model, it is worth noting that echo signals have been widely investigated in other spacetimes, primarily within the ringdown phase of black hole remnants, as well as through the lensing and shadows of ringing black holes, known as lensing tomography \cite{Cardoso:2016rao,Cardoso:2016oxy,Vlachos:2021weq,Chatzifotis:2021pak,Davelaar:2021eoi,Zhong:2024ysg}.}. Specifically for the image, the existence of additional photon sphere produces a multi-ring structure and significantly enhances the flux in the observed intensity. Similar phenomena of multiple photon spheres also occur in scalarized Reissner-Nordstr$\ddot{\text{o}}$m (RN) black holes within the Einstein-Maxwell-scalar (EMS) theory framework. We refer the readers to \cite{Gan:2021pwu,Gan:2021xdl,Guo:2021enm,Guo:2022umh,Guo:2022muy,Chen:2023qic} for more details on the optical appearances and related topics. Beyond these phenomena, alternative photon sphere structures, particularly asymmetric ones, have also been shown to imprint observational signatures in gravitational wave ringdowns, such as spectral instabilities \cite{Cheung:2021bol,Courty:2023rxk,Rosato:2024arw,Destounis:2025dck,Oshita:2024fzf,Spieksma:2024voy,Berti:2022xfj,Ianniccari:2024ysv,Cardoso:2024mrw,Yang:2024vor,Torres:2023nqg,Wang:2025mxe,Jaramillo:2021tmt}.

In this paper, we investigate the interferometric signatures of the hSBH images on long baselines. We particularly focus on how the additional photon sphere influences these signatures and potentially brings in observable deviations from the predictions by GR. By analyzing the structure in visibility amplitudes, we aim to see the prints of multiple photon sphere cases on the interferometric signatures. Previous study \cite{Chen:2023qic} have examined similar interferometric signatures of scalarized RN black hole image in the EMS theory. Compared against such model, the hSBH considered here exhibits a broader generality, because it is constructed via GD approach and does not assume specific matter fields. Thus, this framework indeed provides a flexible platform to study the effects of various types of hair, such as scalar hair, tensor hair, fluid-like dark matter and other components, on the black hole image and its interferometric signatures.

This paper is organized as follows. In Section \ref{review}, we briefly review the hSBH solution, analyze the null geodesic motion of photon outside the black hole, and point out the existence of either a single or double photon spheres. In Section \ref{visibility}, we present the method to obtain the observed intensity of black hole illuminated by the thin accretion disk and their corresponding complex visibility. Specifically, we analytically employ the thin-ring models to analyze the complex visibility and  numerically compute them for hSBHs in Section \ref{analytical} and \ref{numerical}, respectively. We give conclusion and prospect in Section \ref{conclusion}.

\section{A quick review on the photon sphere structure of hairy Schwarzschild black hole}\label{review}

In this section, we will briefly introduce the derivation of hSBH obtained via the GD approach \cite{Ovalle:2020kpd}. Then we analyze the structures of single and double photon spheres outside the black hole by the effective potentials. In this framework, the corresponding Einstein equation is written by
\begin{equation}
G_{\mu\nu}\equiv R_{\mu\nu}-\frac{1}{2}R g_{\mu\nu}=8\pi \tilde{T}_{\mu\nu}, \label{einstein}
\end{equation}
where the total energy momentum tensor $\tilde{T}_{\mu\nu}=T_{\mu\nu}+\vartheta_{\mu\nu}$ include two parts: $T_{\mu\nu}$ the original energy momentum tensor associated with a known solution of GR and $\vartheta_{\mu\nu}$ the introduced energy momentum tensor of new matter fields or additional gravitational sectors. The Bianchi identity requires $\nabla^\mu \tilde{T}_{\mu\nu}=0$. Next we will show main technical hints of GD approach. The spherically symmetric and static solution of Eq.\eqref{einstein} can be expressed by 
\begin{equation}
ds^2=-e^{\nu(r)}dt^2+e^{\lambda(r)}dr^2+r^2(d\theta^2+\sin^2\theta d\phi^2). \label{metric1}
\end{equation}
One can consider that the above solution is assumed to be generated by the following seed metric with the seed source $T_{\mu\nu}$ (i.e. $\vartheta_{\mu\nu}=0$)
\begin{equation}
ds^2=-e^{\xi(r)}dt^2+e^{\mu(r)}dr^2+r^2(d\theta^2+\sin^2\theta d\phi^2).\label{metric2}
\end{equation}
Thus the metric Eq.\eqref{metric1} is attributed to the deformation of seed metric Eq.\eqref{metric2} by introducing the source $\vartheta_{\mu\nu}$, that is 
\begin{equation}
\xi(r)\rightarrow \nu(r)=\xi(r)+\alpha \, k(r),  \qquad e^{-\mu(r)}\rightarrow  e^{-\lambda(r)}=e^{-\mu(r)}+\alpha \, h(r), \label{deform}
\end{equation}
where the parameter $\alpha$ denotes the strength of deformation. Thus, with the above deformations Eq.\eqref{deform}, the Einstein equation Eq.\eqref{einstein} can be split into standard Einstein equation and an additional sector from the deformation, given by
\begin{equation}
{G}_{\mu}^{~\nu}(\xi(r),\mu(r))=8\pi T_{\mu}^{~\nu}, \qquad  \alpha~\mathcal{G}_{\mu}^{~\nu}(\xi(r),\mu(r);k(r),h(r))=8\pi\vartheta_{\mu}^{~\nu}.
\end{equation}
When $\alpha=0$, the solution reduces to the original seed metric as the tensor $\vartheta_{\mu\nu}$ vanishes. The linear decomposition of the Einstein tensor aligns with the linear superposition of sources on the right-hand side of Eq.\eqref{einstein}, which is the key point that makes the GD approach work. The previous steps show that the GD approach enables the systematic construction of deformed solutions by decomposing  the full equation into two decoupled sectors, while the parameter $\alpha$ serves to track the contribution of the new source term $\vartheta_{\mu\nu}$. The GD approach itself is not a perturbative scheme in $\alpha$ and the decoupling is exact and non-perturbative for any value of this parameter \cite{Ovalle:2020kpd}.

Further, considering the SBH solution with vanishing $T_{\mu\nu}$ as the seed metric Eq.\eqref{metric2} and assuming additional source $\vartheta_{\mu\nu}$ as the anisotropic fluid satisfying strong energy condition, one can solve out the Einstein equation and obtain the hairy solution deformed from the SBH metric. Here we omit the derivation steps due to the straightforward nature of the calculations (see \cite{Ovalle:2020kpd} for more details and \cite{Meng:2025ivb} for a liter version) and present the hSBH solution directly
\begin{equation}\label{eq-static}
ds^2=-f(r)dt^2+\frac{dr^2}{f(r)}+r^2(d\theta^2+\sin^2\theta d\phi^2)
~~\mathrm{with}~~ f(r)=1-\frac{2M}{r}+\alpha e^{-r/(M-l_o/2)}, 
\end{equation}
where $M$ is the black hole mass, and $\alpha$ is dimensionless deformation parameter. $l_o=\alpha l$ with $l$ a parameter that has length dimension is the charge of primary hair which should satisfy $l_o \leq 2M$ to guarantee the asymptotic flatness. The event horizon is determined by $f(r_h)=0$, which admits a uniquely positive root. Note that this metric characterizes specific deformations of the SBH solution that arise from the introduction of extra material sources. These sources may include scalar hair, tensor hair, fluid-like dark matter, and other components. When the deformation parameter $\alpha=0$, the metric Eq.\eqref{eq-static} reduces to the standard SBH solution in GR, which corresponds to the case without matter source. Throughout the paper, all the physical quantities are in units of the black hole mass by setting $M=1$ for simplicity and without loss of generality.

Next we will analyze the structures of photon spheres by building equations of null geodesic motion outside the hSBH. The Lagrangian of photon is given by
\begin{equation}
\mathcal{L}=\frac{1}{2}g_{\mu\nu}\dot{x}^\mu\dot{x}^\nu=\frac{1}{2}\left(-f(r)\dot{t}^2+\frac{1}{f(r)}\dot{r}^2+r^2\left(\dot{\theta}^2+\text{sin}^2\theta \dot{\phi}^2\right)\right),\label{lag}
\end{equation}
where $\dot{x}^\mu=dx^\mu/d\lambda$ represents the four-velocity of photon and $\lambda$ is the affine parameter. Due to the spherical symmetry of the spacetime, we focus on the photons moving on the equatorial plane ($\theta=\pi/2$) for convenience. Considering $\mathcal{L}=0$ for the photon and the following Euler-Lagrange equation that determines their motions
\begin{equation}\label{eq-ELeq}
\frac{d}{d\lambda}\left(\frac{\partial \mathcal{L}}{\partial \dot{x}^\mu}\right)=\frac{\partial \mathcal{L}}{\partial x^\mu},
\end{equation}
one can finally derive the radial component of null geodesic equation
\begin{align}
&\dot{r}^2=\frac{1}{b^2}-V_{\text{eff}}(r)~~\mathrm{with}~~V_{\text{eff}}(r)=\frac{f(r)}{r^2}, \label{radialEq}
\end{align}
where $V_{\text{eff}}$ is the effective potential, $b\equiv L_z/E$ is the impact parameter, and the affine parameter $\lambda$ is redefined as $\lambda/L_z$. Note that $L_z$ and $E$ denote the conserved $z$-component of angular momentum and the energy for photons, respectively. We are interested in the unstable photon sphere, where the radius $r_{\text{ph}}$ and critical impact parameter $b_{\text{ph}}$ of photon sphere are determined by 
\begin{equation}
V_{\text{eff}}(r_{\text{ph}})=\frac{1}{b_{\text{ph}}^2}, \qquad V_{\text{eff}}'(r_{\text{ph}})=0, \qquad V_{\text{eff}}''(r_{\text{ph}})<0. \label{formulabrph}
\end{equation}
Thus, the unstable photon spheres are indicated by the local maximum of the effective potential. Given that only unstable photon spheres determine the properties of black hole image \cite{Perlick:2021aok}, we will focus on these unstable photon spheres in the following.

\begin{figure}[htbp]
\centering
{\includegraphics[width=5.5cm]{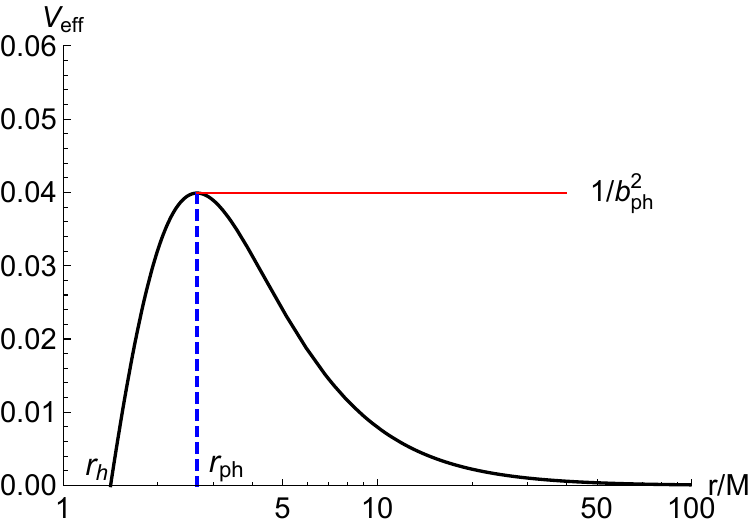}}\hspace{5mm}
{\includegraphics[width=5.5cm]{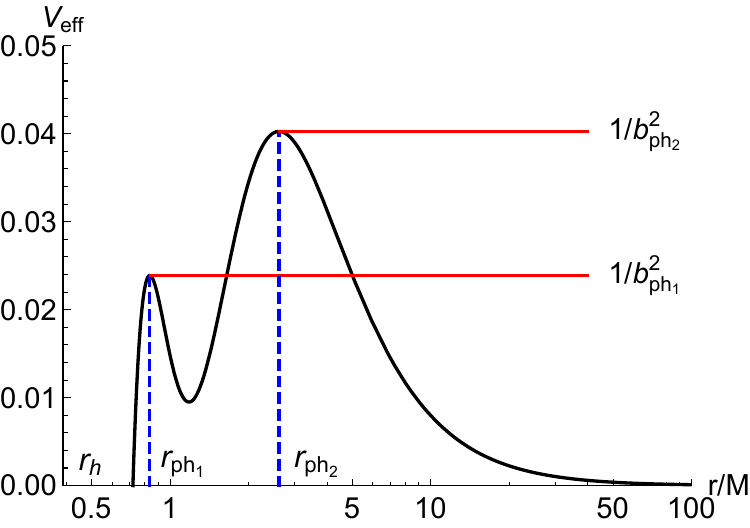}}\hspace{5mm}
{\includegraphics[width=5.5cm]{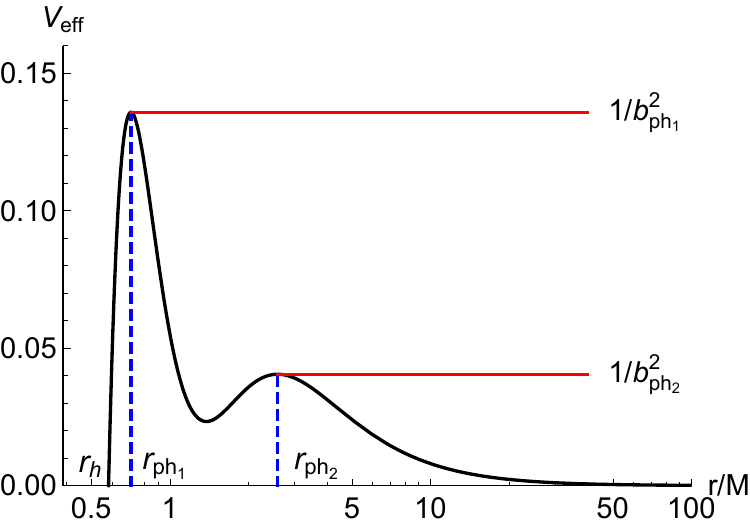}}
\caption{The effective potential $V_{\text{eff}}$ of the hSBH with single (\textbf{Left}) and double (\textbf{Middle} and \textbf{Right}) photon spheres. We choose $\alpha=7$, $\alpha=7.5$ and $\alpha=7.8$, respectively, with fixed $l_o=1$ and $M=1$.}
\label{figVeff}
\end{figure}

Unlike the standard SBH that possesses a single photon sphere, the hSBH can present double photon spheres in some certain parameters \cite{Guo:2022ghl}, inspired by which, we carefully studied the structures of photon spheres in our previous work \cite{Meng:2023htc}. In terms of the photon sphere features, we divided the parameter regime into three parts in which the corresponding effective potential has three typical configurations. Here we fix $l_o=1$ and tune the deviation parameter to explicitly show all three typical configurations in Fig.\ref{figVeff}, from which one can directly read off the photon sphere structures from the feature of the effective potentials. The left panel with $\alpha=7$ exhibits a single photon sphere, similar to that for SBH. The middle and right panels show that for $\alpha=7.5$, the inner peak is lower than the outer one, while for $\alpha=7.8$, the inner peak is higher. Notably, the two local maxima in the effective potential enable the hSBH to admit double photon spheres.

In our previous work \cite{Meng:2023htc}, we also investigated the optical appearances of the hSBH illuminated by thin accretion disks. The results showed that, for hSBH possessing double photon spheres, with the potential's inner peak higher than the outer one, the images exhibit additional bright rings and distinctive accretion features, which are absent in the case with a single photon sphere. It is worth noting that there also exists a scenario in which the inner peak is lower than the outer one, as shown in the middle panel of Fig.\ref{figVeff}. However, we have addressed in \cite{Meng:2023htc} that, in such case both the light rays distribution and the image  are similar to those in the case with a single photon sphere, because the photons trapped near the lower inner photon sphere cannot escape the black hole's gravity and thus remain unobservable to a distant observer. This point will soon be verified via the interferometric signatures.

\section{Observed intensity and complex visibility} \label{visibility}

In this section, we will explore the interferometric signatures for the hSBH images with single and double photon spheres on long interferometric baselines. To proceed, we shall discuss the images of black hole illuminated by the thin accretion disk that is assumed to emit isotropically in the rest frame of static worldlines, as motivated by the fact that hot, optically thin accretion flows are surrounding M87*, Sgr A* and many other supermassive black holes in our Universe \cite{Yuan:2014gma}. We assume that the observer is located on the disk’s axis of symmetry and thus views the accretion disk face-on. The thin accretion disk considered herein is simple but enough for the purpose of this paper.

The observed intensity $I_{\text{obs}}$ of black hole image is given by \cite{Gralla:2019xty}
\begin{equation}
I_{\text{obs}}(b)=f(r)^2 I_{\text{em}}(r)=\sum_{m}f(r)^2I_{\text{em}}(r)|_{r=r_m (b)}, \label{eqintensity}
\end{equation}
where $I_{\text{em}}$ represents the emitted intensity, and the transfer function $r_m (b)$ denotes the radial coordinate $r$ of $m$-th intersection between the light ray with impact parameter $b$ and the accretion disk. Since the light ray extracts energy from the accretion disk during each intersection, different types of light rays will contribute differently to the observed intensity. By the definition proposed in \cite{Gralla:2019xty,Wang:2023vcv}, one can classify the light rays into three types based on the intersection times. Concretely, light rays with $m=1$, $m=2$ and $m\geq 3$ are regarded as the direct, lensed ring and photon ring intensity of black hole images, respectively. As illustrated in \cite{Johnson:2019ljv}, GR predicts that the black hole image contains the structure of a thin photon ring consisting of an infinite sequence of self-similar subrings characterized by the index $m$. As $m$ increases, these subrings asymptotically approach the edge of black hole shadow and finally form the photon sphere in the limit $m\rightarrow \infty$. Note that the observed intensity Eq.\eqref{eqintensity} is a function of impact parameter $b$. In the following, we will introduce $\beta=b/r_o$ for convenience, where $r_o$ is the distance from the center of black hole to distant observer and $\beta$ denotes the angle between the incident ray and axis defined by the connecting line between the center of black hole and the observer.

Given the computational method Eq.\eqref{eqintensity} of black hole image, we will introduce the emitted intensity of accretion disk called Gralla-Lupsasca-Marrone (GLM) model, which is shown that it is close to match the observational predictions for the emission profiles of astrophysical accretion disks obtained by general relativistic magnetohydrodynamics simulations \cite{Vincent:2022fwj}. The emission function is the following \cite{Gralla:2020srx}
\begin{equation}
I_{\text{em}}(r)=\frac{\text{exp}\left\{-\frac{1}{2}\left[\gamma+\text{arcsinh}\left(\frac{r-\mu}{\sigma}\right)\right]^2\right\}}{\sqrt{(r-\mu)^2+\sigma^2}}, \label{profile}
\end{equation}
where the shape of emission profile is controlled by the free parameters $\gamma$, $\mu$, and $\sigma$. Concretely, $\gamma$ controls the rate of increase of the emission profile from infinity down to the peak, $\mu$ performs a translation of emission profile, and $\sigma$ controls the dilation of emission profile. This model also has been used to investigate observational appearances of BHs and other ultracompact objects \cite{Rosa:2023hfm,Rosa:2023qcv,Gao:2023mjb,Rosa:2024bqv}. For our purpose, we set the parameters as $\gamma=-3/2$, $\mu=0$ and $\sigma=M/2$, which are also chosen in \cite{Gralla:2020srx}.

In the above, we have provided the method to obtain the black hole image. However, interferometric arrays, such as the EHT and next generation \cite{Johnson:2023ynn,Ayzenberg:2023hfw,Doeleman:2023kzg} or space-based telescopes \cite{Lupsasca:2024xhq,Galison:2024bop,Johnson:2024ttr}, do not directly sample a black hole image, but rather radio complex visibility$-$the Fourier transform of the observed intensity \cite{Thompson2017Book},
\begin{equation}
V(\boldsymbol{u})=\int I_{\text{obs}}(\boldsymbol{x})e^{-2\pi i \boldsymbol{u}\cdot \boldsymbol{x}}d^2\boldsymbol{x}, \label{complexV1}
\end{equation}
where $\boldsymbol{u}$ is dimensionless baseline vector that corresponds to the distance between telescopes in the array,  projected onto the plane perpendicular to the line of sight and measured in units of the observation wavelength. $\boldsymbol{x}$ is the angular coordinates on the image plane. Due to the axisymmetric image in our case Eq.\eqref{eqintensity}, the complex visibility Eq.\eqref{complexV1} can be simplified into the zero-order Hankel transform \cite{Cardenas-Avendano:2023dzo}
\begin{equation}
V(u)=2\pi \int I_{\text{obs}}(\beta)J_0(2\pi\beta u)\beta d \beta, \label{complexV2}
\end{equation}
where $u=|\boldsymbol{u}|$ is the baseline length and $J_0$ denotes the zero-order Bessel function of the first kind.

Next we shall first adopt the analytical method proposed in \cite{Chen:2023qic} to study how photon sphere structure affect the interferometric signatures of black hole image by thin-ring models, and then numerically illustrate these characteristics in hSBH.

\subsection{Analytical hints}\label{analytical}

We firstly consider the black hole with single photon sphere at the radius of critical impact parameter denoted by $b_{\text{ph}}$. Because the width of the photon ring intensity is significantly narrower in comparison to that of the direct and lensed ring intensities \cite{Gralla:2019xty}, the $\delta$-function can be used to represent an infinitesimally thin, uniform and circular ring as photon ring intensity approximately with the peak located at the angular radius $\beta_{\text{ph}}=b_{\text{ph}}/r_{o}$, which is expressed as \cite{Johnson:2019ljv} 
\begin{equation}
I_{\text{obs}}^{\text{ph}}(\beta)=\Phi_{\text{ph}}\delta(\beta-\beta_{\text{ph}}),
\end{equation}
where $\Phi_{\text{ph}}$ is a constant. We refer to this as the single-thin-ring model. Further, the complex visibility Eq.\eqref{complexV2} is written as
\begin{equation}
V(u)=2\pi \Phi_{\text{ph}}\beta_{\text{ph}}J_0(2\pi \beta_{\text{ph}} u).
\end{equation}
For lone baselines, the complex visibility admits the asymptotic expansion
\begin{equation}
V(u)\approx 2 \Phi_{\text{ph}}\sqrt{\beta_{\text{ph}}}\frac{\text{cos}(2\pi\beta_{\text{ph}}u-\pi/4)}{\sqrt{u}}.
\end{equation}
Obviously, $V(u)$ exhibits the weakly damped oscillation behavior with the period $1/\beta_{\text{ph}}$ decaying as $1/\sqrt{u}$.

For the black hole with double photon spheres, if the peak of effective potential at inner photon sphere is lower than the outer one, the inner photon sphere could slightly affect the observed intensity because the nearby photons cannot escape; however, if the peak at inner photon sphere is higher than the outer one, both photon spheres non-trivially influence the observational feature  \cite{Gan:2021xdl,Meng:2023htc}. Consequently, the complex visibility for the former double photon sphere structure should also have a damped oscillation behavior, similar to the case with a single photon sphere. For the latter double photon sphere structure, we denote the angular radii of critical impact parameters for the inner and outer photon spheres as $\beta_{\text{ph}}^{\text{in}}$ and $\beta_{\text{ph}}^{\text{out}}$, respectively. Then the photon ring intensity can be represented by two infinitesimally thin, uniform and circular rings approximately,
\begin{equation}
I_{\text{obs}}^{\text{ph}}(\beta)=\Phi_{\text{in}}\delta(\beta-\beta_{\text{ph}}^{\text{in}})+\Phi_{\text{out}}\delta(\beta-\beta_{\text{ph}}^{\text{out}}),\label{model2}
\end{equation}
where $\Phi_{\text{in}}$ and $\Phi_{\text{out}}$ are constants. We refer to this as the double-thin-ring model. Subsequently, the complex visibility is 
\begin{equation}
V(u)=2\pi \Phi_{\text{in}}\beta_{\text{ph}}^{\text{in}}J_0(2\pi \beta_{\text{ph}}^{\text{in}} u)+2\pi \Phi_{\text{out}}\beta_{\text{ph}}^{\text{out}}J_0(2\pi \beta_{\text{ph}}^{\text{out}} u). 
\end{equation}
For lone baselines, the above formula becomes
\begin{equation}
V(u)\approx 2 \Phi_{\text{in}}\sqrt{\beta_{\text{ph}}^{\text{in}}}\frac{\text{cos}(2\pi\beta_{\text{ph}}^{\text{in}}u-\pi/4)}{\sqrt{u}}+2 \Phi_{\text{out}}\sqrt{\beta_{\text{ph}}^{\text{out}}}\frac{\text{cos}(2\pi\beta_{\text{ph}}^{\text{out}}u-\pi/4)}{\sqrt{u}}.
\end{equation}
In contrast to the scenarios involving a single photon sphere as well as the former double photon sphere structure, the complex visibility for the latter double photon sphere structure displays not only the damped oscillation behavior, but also beat signals with the period $\Delta u=1/(\beta_{\text{ph}}^{\text{out}}-\beta_{\text{ph}}^{\text{in}})$. These findings will be numerically verified in the next subsection.

\subsection{Numerical results} \label{numerical}
To figure out the complete behavior of complex visibility, we will numerically compute the optical appearances of hSBH with the three typical photon sphere structures (see Fig.\ref{figVeff}), and their corresponding interferometric signatures, respectively. As defined in the previous section, the interferometric signature of image is the Fourier transform of the source's observed intensity on the sky, known as the complex visibility. Our analysis will focus on this quantity to disclose how the presence of multiple photon spheres affects the interferometric signatures. We assume that the accretion disk, which is located on the equatorial plane of the black hole, is viewed face-on by a distant observer at $r_o=100$. Here it is enough to use the GLM emission Eq.\eqref{profile} to present interferometric characteristics of images for multiple photon spheres.

\begin{figure}[htbp]
\centering
{\includegraphics[width=5.6cm]{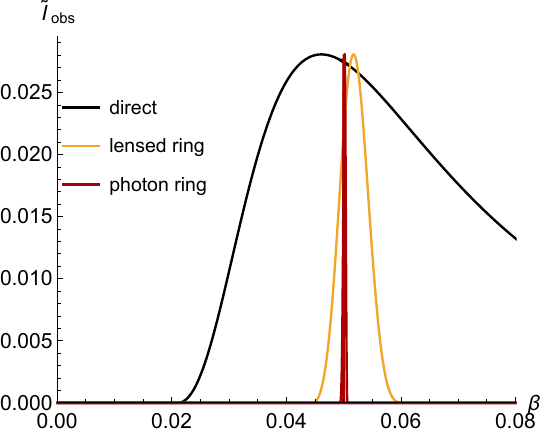}}\hspace{3mm}
{\includegraphics[width=5.6cm]{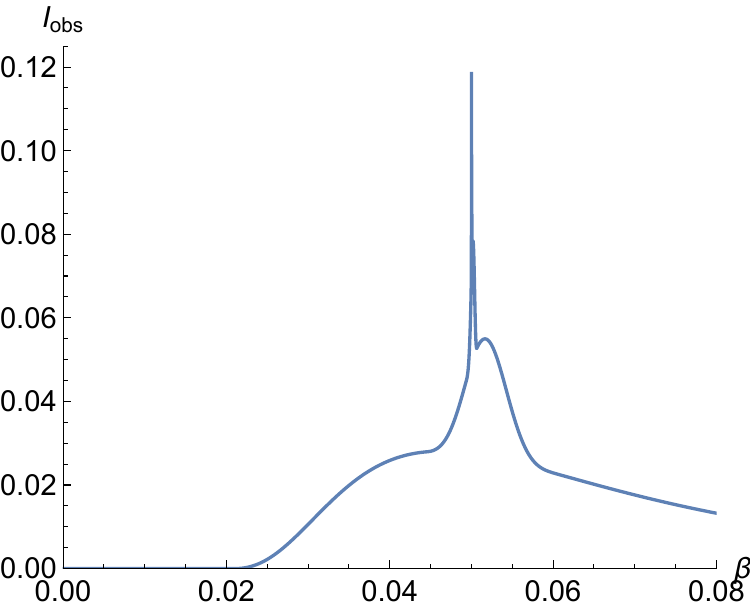}}\hspace{3mm}
{\includegraphics[width=4.5cm]{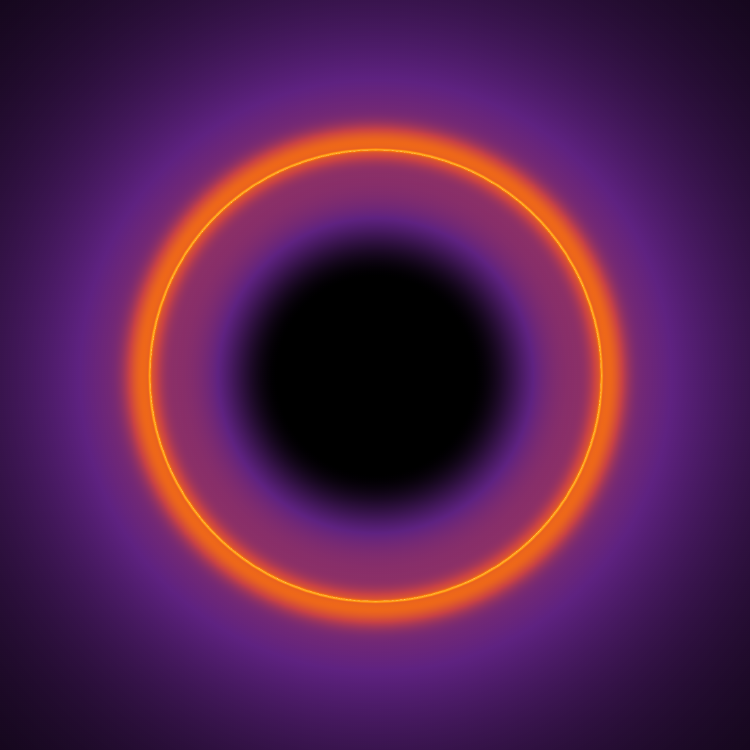}}\vspace{5mm}
{\includegraphics[width=11cm]{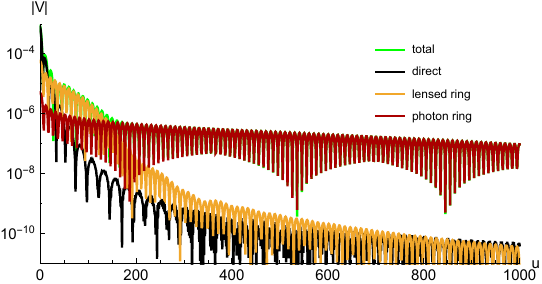}}
\caption{The hSBH with a single photon sphere. Here we fixed $\alpha=7$, $l_o=1$ and $M=1$, with the effective potential shown in the left panel of Fig.\ref{figVeff}. The angle of photon sphere is $\beta_{\text{ph}}=0.05$. \textbf{Top-left}: different observed intensities originated from the direct (black), lensed ring (gold) and photon ring (red) intensities, respectively. \textbf{Top-middle}: the observed intensities as a function of angle $\beta$. \textbf{Top-right}: optical appearance: the intensity distribution across a two-dimensional plane. \textbf{Bottom}: total visibility amplitude $|V(u)|$ (green) and its components contributed by direct (black), lensed ring (gold) and photon ring (red) intensities, respectively.} 
\label{figSingle}
\end{figure}

We start with exploring the single photon sphere, for which the effective potential is shown in the left panel of Fig.\ref{figVeff}. The critical impact parameter is $b_{\text{ph}}=5.01$ with the corresponding angle $\beta_{\text{ph}}=0.05$.
The individual contributions of the direct (black), lensed ring (gold), and photon ring (red) intensities, along with the observed intensity $I_{\text{obs}}(\beta)$, are shown in the top-left and top-middle panels of Fig.\ref{figSingle}, respectively. Obviously, it displays a sharp peak at a radius close to that of photon sphere. The top-right panel of Fig.\ref{figSingle} shows the corresponding black hole image. However due to the narrow width of photon ring intensity, it makes the negligible contribution to the black hole image.

The earlier studies have demonstrated that any smooth ring with width $w$ can be resolved and the visibility of such rings decay exponentially provided that the interferometric baseline is sufficiently long $(u\gg 1/w)$ \cite{Johnson:2019ljv}. This indicates that as the width $w$ of ring decrease, the longer baseline $u$ is required to resolve its structure because this certain ring dominates the complex visibility signal. To better understand the above description, we compute the corresponding complex visibility amplitudes $|V(u)|$ from observed intensities by using Eq.\eqref{complexV2} and the results are shown in the bottom panel of Fig.\ref{figSingle}. We find that as the baseline length increases, the direct, lensed ring and photon ring intensities become dominant in turn. Specifically, at long baseline, the complex visibility of photon ring intensity exhibits the damped oscillation as previously discussed.

\begin{figure}[htbp]
\centering
{\includegraphics[width=5.6cm]{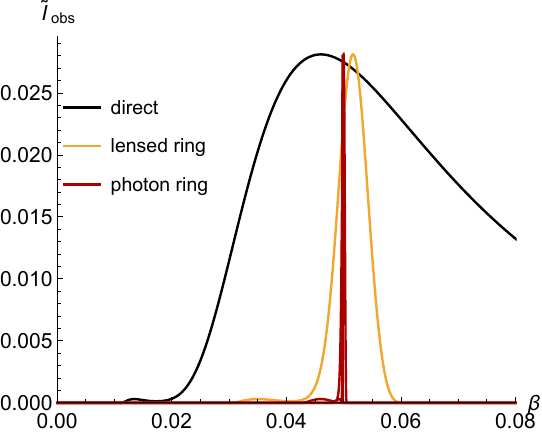}}\hspace{3mm}
{\includegraphics[width=5.6cm]{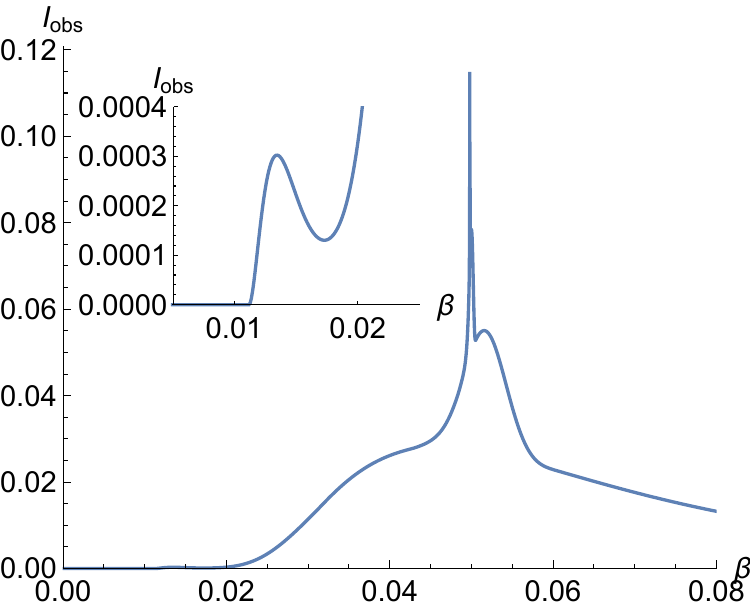}}\hspace{3mm}
{\includegraphics[width=4.5cm]{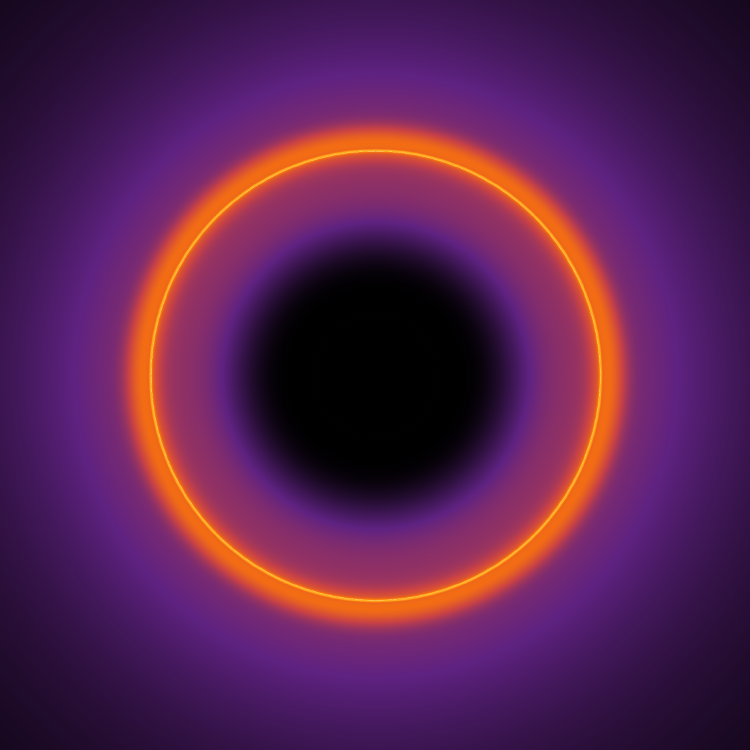}}\vspace{5mm}
{\includegraphics[width=11cm]{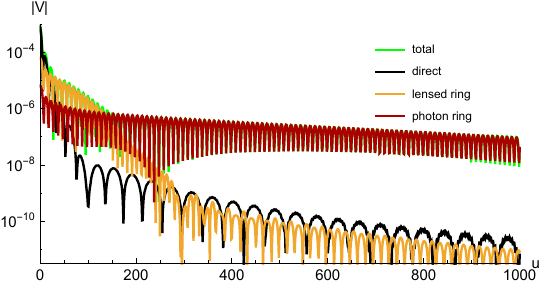}}
\caption{The hSBH with double photon spheres. Here we fixed $\alpha=7.5$, $l_o=1$ and $M=1$, with the effective potential shown in the middle panel of Fig.\ref{figVeff}. \textbf{Top-left}: different observed intensities originated from the direct (black), lensed ring (gold) and photon ring (red) intensities, respectively. \textbf{Top-middle}: the observed intensities as a function of angle $\beta$. \textbf{Top-right}: optical appearance: the intensity distribution across a two-dimensional plane. \textbf{Bottom}: total visibility amplitude $|V(u)|$ (green) and its components contributed by direct (black), lensed ring (gold) and photon ring (red) intensities, respectively.}
\label{figDoubleNew}
\end{figure}

Then we study the double photon spheres with a lower inner peak, where their effective potential shown in the middle panel of Fig.\ref{figVeff}. The associated observed intensity and visibility amplitude are depicted in Fig.\ref{figDoubleNew}, from which we observe two expected properties as we addressed in the analytical subsection. On one hand, despite the presence of extremely low brightness in the small impact parameter region, the observed intensity is comparable to that of the single photon sphere case. On the other hand, its complex visibility also resembles that of the single photon sphere case.

\begin{figure}[htbp]
\centering
{\includegraphics[width=5.6cm]{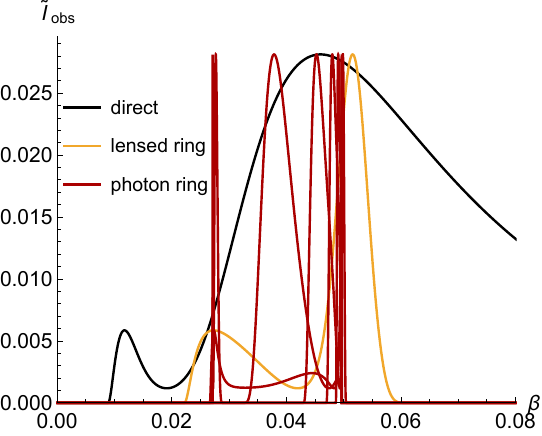}}\hspace{3mm}
{\includegraphics[width=5.6cm]{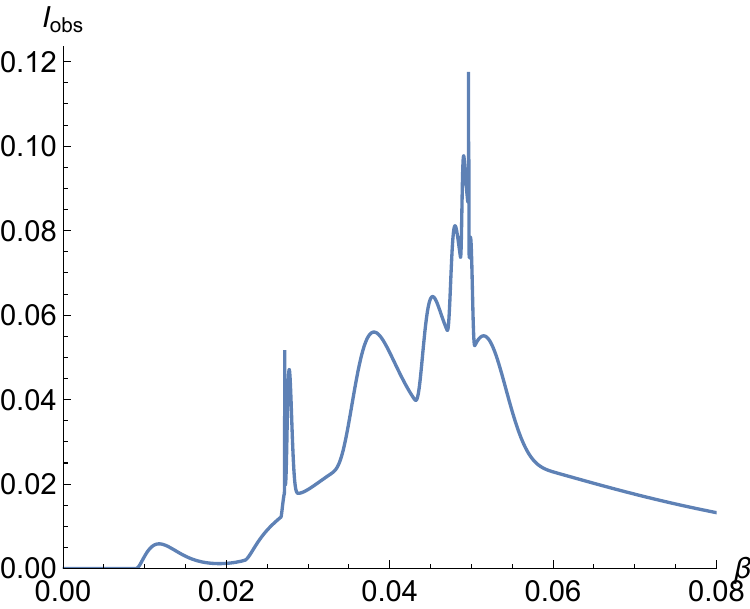}}\hspace{3mm}
{\includegraphics[width=4.5cm]{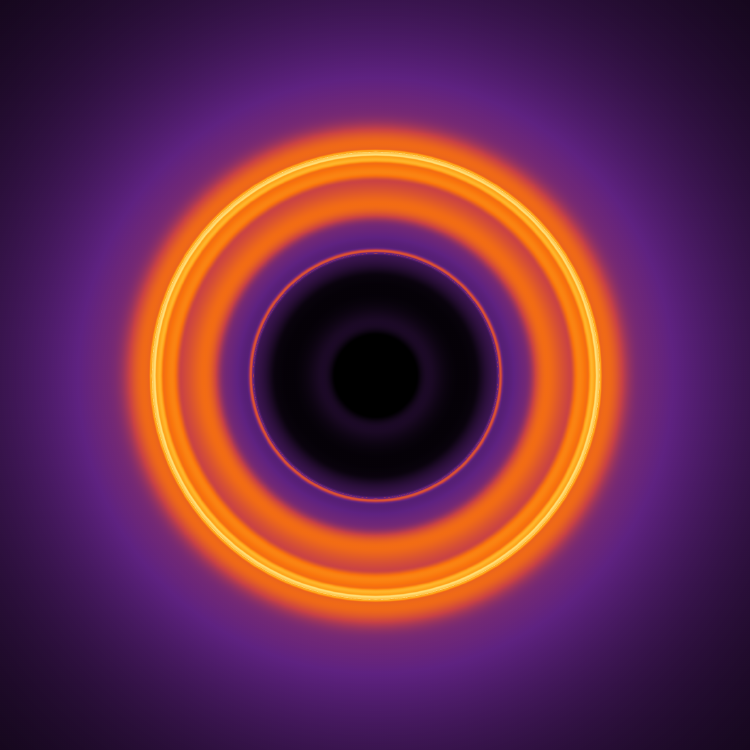}}\vspace{5mm}
{\includegraphics[width=11cm]{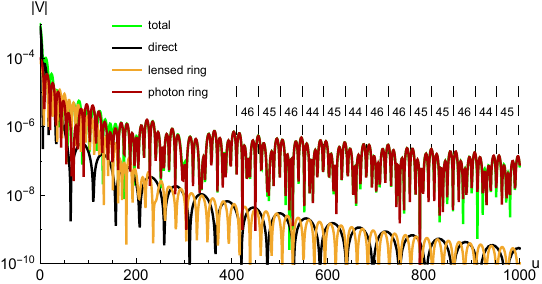}}
\caption{The hSBH with double photon spheres. Here we fixed $\alpha=7.8$, $l_o=1$ and $M=1$, with the effective potential shown in the right panel of Fig.\ref{figVeff}. The angles of double photon spheres are $\beta_{\text{ph}}^{\text{in}}=0.0271$ and  $\beta_{\text{ph}}^{\text{out}}=0.0497$. \textbf{Top-left}: different observed intensities originated from the direct (black), lensed ring (gold) and photon ring (red) intensities, respectively. \textbf{Top-middle}: the observed intensities as a function of angle $\beta$. \textbf{Top-right}: optical appearance: the intensity distribution across a two-dimensional plane. \textbf{Bottom}: total visibility amplitude $|V(u)|$ (green) and its components contributed by direct (black), lensed ring (gold) and photon ring (red) intensities, respectively. For the sufficiently long baseline, the beat pattern will emerge. The widths of two adjacent beat patterns match well with $1/(\beta_{\text{ph}}^{\text{out}}-\beta_{\text{ph}}^{\text{in}})=44.25$.}
\label{figDouble}
\end{figure}

Next we move on to analyze the double photon spheres with a higher inner peak, where their effective potential is presented in the right panel of Fig.\ref{figVeff}. The critical impact parameters of inner and outer peaks are $b_{\text{ph}}^{\text{in}}=2.71$ and $b_{\text{ph}}^{\text{out}}=4.97$, with the corresponding angles $\beta_{\text{ph}}^{\text{in}}=0.0271$ and $\beta_{\text{ph}}^{\text{out}}=0.0497$, respectively. The top-left panel of Fig.\ref{figDouble} displays individual contributions of the direct (black), lensed ring (gold), and photon ring (red) intensities. The corresponding observed intensity $I_{\text{obs}}(\beta)$ is presented in the top-middle panel. Compared to the previous two cases, this double photon sphere structure broadens the width of photon ring, significantly enhancing its contribution to the observed intensity. Meanwhile it also results in the structures with multiple peaks due to the complex trajectories of light rays. The corresponding black hole image depicted in top-right panel of Fig.\ref{figDouble} exhibits the multiple rings that demonstrates significant contribution to the observed intensity from photon ring intensity.

The bottom panel of Fig.\ref{figDouble} shows the complex visibility amplitudes $|V(u)|$ for both corresponding observed intensity and individual contributions, respectively. Obviously, the visibilities of direct and lensed ring intensities are similar to those observed in both the single photon sphere and the former double photon sphere structures. However, in the case of photon ring intensity, the visibility displays intricate structures that are markedly distinct from the interferometric signatures of previous two cases. This difference arises from the existence of double photon spheres with the inner one corresponding to higher peak in photon's effective potential. Concretely, at short baseline, the direct intensity component dominates the visibility signal, while both the lensed and photon ring contributions remain unresolved. As the baseline length increases, the visibility of photon ring will eventually dominate the total visibility. Specially, the visibility signal exhibits the beat pattern with the period approximately described by $\Delta u=1/(\beta_{\text{ph}}^{\text{out}}-\beta_{\text{ph}}^{\text{in}})=44.25$, which can be well approximated by the double-thin-ring model Eq.\eqref{model2} mentioned in the analytical analysis.

\section{Conclusion and discussion}\label{conclusion}

The photon ring in the black hole image produces strong and universal interferometric signatures on long baselines. These signatures offer the possibility of precision measurements of black hole parameters and tests of GR. Moreover, the hSBH constructed by introducing additional sources exhibits great generality because there are no certain matter fields in the GD approach. It allows us to study the black hole image and interferometric signature effected by arbitrary type of hair (e.g. scalar hair, tensor hair, fluid-like dark matter, etc). Specifically, the hairy metric can have double photon spheres for certain parameter regions. Thus we can explore how the additional photon sphere impacts the interferometric characteristics.

We began with a brief review on hSBH, and illustrated that three types of configurations in the effective potentials for photons correspond to three photon sphere structures. Then we presented the computational method employed to generate image of hSBHs illuminated by the thin accretion disk. Furthermore, by adopting the simplified model where the photon ring intensity in the image is represented by single or double infinitesimally thin, uniform, circular rings, we derived analytical expressions for the complex visibility. Subsequently, we numerically computed the observed intensities and images of hSBHs featuring both single and double photon spheres, along with their corresponding complex visibilities.

Our results for the single photon sphere case reveal a damped oscillatory behavior in the visibility amplitude. Meanwhile, for the case of double photon spheres with a lower inner peak, the results are similar to the single photon sphere. However, when the inner peak is higher than the outer peak for double photon spheres, the results exhibit significantly departure from the cases with other photon sphere structures. The photon ring makes the non-negligible contribution to the observed intensity. Moreover, the visibility signal displays a beat pattern, the period of which can be well estimated using a double-thin-ring model. These distinct features could produce potentially detectable signatures with future detectors.

The current studies could disclose the potentially universal interferometric signatures that the black holes with double unstable photon spheres display beat signals. However, the stable photon sphere, which is located between two unstable photon spheres, plays a central role in the instability mechanism of black holes and exotic compact objects. The study in \cite{Guo:2021enm,Yang:2025hqk} indicate that some quasi-normal modes can be trapped around this stable region for a long time. The accumulation of these long-lived modes may eventually develop into a nonlinear instability \cite{Keir:2014oka,Cardoso:2014sna}. Recently, fully nonlinear numerical evolutions have confirmed that the stable photon sphere in ultracompact objects can trigger this instability \cite{Cunha:2022gde}. But the author of \cite{Guo:2024cts} demonstrated that scalarized RN black holes in EMS theory exhibit nonlinear stability under spherically symmetric scalar perturbations, despite the presence of a stable photon sphere. Thus, it is essential to perform full nonlinear numerical evolution of our hSBH spacetime in future research. Moreover, these results in our work are based on assumptions that the black hole is illuminated by thin accretion disk, viewed face-on. To strengthen the astrophysical relevance and universality of these findings, several directions deserve further exploration.  Firstly, the accretion disk models employed here are ideal. Incorporating more realistic accretion flows, and particularly those surround the rotating counterpart of the black hole, should model more realistic astrophysical environment.  Moreover, the line of sight for the observer is usually offset from the disk's axis. Thus, we could examine how the inclination angle of observers affects our findings, particularly whether it influences interferometric signatures and the beat period for the case of double photon sphere. From an observational aspect, our results could provide theoretical templates for high-resolution interferometric facilities such as the EHT \cite{EventHorizonTelescope:2019dse} and the next-generation EHT \cite{Johnson:2023ynn}. Moreover, it is interesting to study the connections between these beat signals and the black hole glimmer signatures \cite{Wong:2020ziu,Hadar:2023kau,Wong:2024gph} as well as the black hole flares \cite{Broderick:2005my}, which are the mid-range observational goal for the EHT \cite{EventHorizonTelescope:2024whi}.

\begin{acknowledgments}
This work is supported by the National Natural Science Foundation of China under Grants Nos. 12447137, 12222302 and 12375054.
\end{acknowledgments}

%\newpage

%\bibliographystyle{JHEP}
\bibliographystyle{utphys}
\bibliography{ref}

\end{document}